\documentclass{article} 
\pdfoutput=1

\usepackage{nips12submit_e,times}
\usepackage{amsfonts}
\usepackage{graphicx}
\usepackage{cite}

\title{Deep learning and the renormalization group}

\author{
C\'edric B\'eny\\
Institut f\"ur Theoretische Physik\\
Leibniz Universit\"at Hannover\\
Appelstra{\ss}e 2, 30167 Hannover, Germany\\
\texttt{cedric.beny@gmail.com}
}

\nipsfinalcopy 

\begin{document}

\maketitle

\begin{abstract}
Renormalization group (RG) methods, which model the way in which the effective behavior of a system depends on the scale at which it is observed, are key to modern condensed-matter theory and particle physics.
We compare the ideas behind the RG on the one hand and deep machine learning on the other, where depth and scale play a similar role. In order to illustrate this connection, we review a recent numerical method based on the RG---the multiscale entanglement renormalization ansatz (MERA)---and show how it can be converted into a learning algorithm based on a generative hierarchical Bayesian network model. Under the assumption---common in physics---that the distribution to be learned is fully characterized by local correlations, this algorithm involves only explicit evaluation of probabilities, hence doing away with sampling. 
\end{abstract} 

\maketitle

Natural sciences extend the process by which we intuitively discover patterns in our sensory data to domains beyond our natural sensory abilities, but also beyond our intuitive reasoning abilities. 
Already, machine learning methods are becoming important in fields where the patterns are too complex to be modeled with simple equations, and too far removed from intuition to be comprehended without formal tools.   

Physics may seem an unlikely candidate for application of machine learning, given that it specifically focuses on systems which can be understood with simple laws. Nevertheless, these very compact laws have to be compared to experiments, and hence require a ``decompression'' process which rapidly leads to overwhelming levels of complexity.

The {\em renormalization group} (RG) was designed to handle some of these complexities 
and had a tremendous influence in particle physics and condensed-matter physics~\cite{wilson83}. It models the dependence of a system's effective behavior on a parameter which is usually thought of as {\em scale} or {\em energy}, but which we will think of here as {\em depth}. Although the RG idea is not recent, new approaches have emerged which have the potential to fully automatize it~\cite{vidal07,vidal08,perez06,verstraete06}. These methods are designed for simulation rather than learning, but, because they are defined through clever representations of certain classes of (quantum) states, they can be adapted to learning tasks as well~\cite{cramer11,landoncardinal12}.

In this paper, I will explain how one of these methods (the multiscale entanglement renormalization ansatz~\cite{vidal09}) can be made into a deep learning algorithm for classical probability distributions, assuming no prior knowledge of RG or MERA, nor of quantum mechanics. The machine learning framework task that I am considering is that of building a representation of a probability distribution which optimizes the likelihood of the training data.

\section{The renormalization group}
\label{RG}

Let us take the point of view of condensed matter physics, where the system of interest is a $d$-dimensional {\em lattice} which we think of as a graph $\Gamma$ which could be embedded in $\mathbb R^d$ in such a way that the edges loosely indicate proximity of sites. Hence we can think of it as a discrete version of a $d$-dimensional space. We take the graph as fixed once and for all, but we associate with each vertex $v$, which we call a {\em site}, a random variable $X_v$ which can take any of $n$ possible values. The joint probability distribution for all these variables will be referred to as the lattice's {\em state}, and is the object that we want to model. 

In the context of machine learning, an example would be a probability distribution over images. In this example, the dimension is $d=2$, the sites are pixels linked to their four neighbors, and the random variables are the pixels color values.

Physicists are often interested in such system at thermal equilibrium, where the state can be encoded in a ridiculously compact way through the system's Hamiltonian, or energy function, which is proportional to the logarithm of the probability distribution. What makes this description compact is that the Hamiltonian is usually a sum of local functions, i.e., functions which depend only on a little group of neighboring sites. Hence, the state is a Markov random field with respect to the graph $\Gamma$. 

However, deducing any property of the state from the Hamiltonian (such as marginal probabilities) is generally very difficult. In fact, the large majority of the work done in theoretical physics is purely concerned with doing just that: deriving approximate properties of systems whose state is defined by a given Hamiltonian, i.e., decompressing the Hamiltonian description, so that predictions can be compared to experiments. In this sense, the Hamiltonian, although it is a compact representation of the state, is not very useful as a representation of what one knows about the system's properties.

A Hamiltonian is not, however, entirely useless. Given a Hamiltonian, it is relatively easy to make predictions about short-range correlations. For instance, by sampling the distribution using a Monte-Carlo simulation. However, the computational cost grows exponentially with the range of the correlations. 

The RG approach is based on the observation that longer range correlations can often be summarized by an {\em effective} Hamiltonian on a smaller array of coarse-grained random variables which are, for instance, averages of neighboring sites. That is, one can build a stochastic map from the state of the system to that of coarser array of new variables in such a way that the new state can also be described by a local Hamiltonian.
This process can be iterated, leading to a hierarchy of effective Hamiltonians labeled by a scale. This family, if it could be computed, would provide an efficient encoding of the system's behavior at all scales.

The actual choice of ``binning'', i.e., the form of the stochastic maps which implement the coarse-grainings, is an essential element of the process, as it selects the proper, emergent, {\em order parameters} which contain non-trivial information about the large-scale properties of the system. One may think of these local order parameters as higher-order concepts needed to understand the full state. Typically, the only variables left at the largest scale are the thermodynamic variables which are averages over the whole system. If the coarse-graining procedure is chosen correctly, those variables should identify in which thermodynamic phase the system finds itself. 

This hierarchical scale-by-scale encoding of a probability distribution resembles, for instance, a deep encoding of an image whose features are defined by iteratively combining locale patterns into more complex---and larger---patterns. 
Where this analogy fails, however, is that the distribution to be learned in the context of image recognition is likely not generated by a local Hamiltonian. Nevertheless, we may still think of a local Hamiltonian which appropriately reproduces the short-range correlations within neighboring pixels. For instance, this Hamiltonian could assign higher probabilities to edge filters. The edge orientations themselves may then be the next coarser variables governed by their own effective Hamiltonian which prefers certain assembly of neighboring edges, and so on until, at the deepest level, the only variables left encodes what the whole image represents, with an effective Hamiltonian yielding the relative frequency of the different classes in the learning data.

The renormalization ``group'' denotes the set of coarse-graining operations with composition as group operation, although it is rarely a group given that its elements typically have no inverse, and cannot all be composed together. It is more fruitful to think of it as a flow on the ``manifold'' of Hamiltonians (assuming a continuum of scales), where each possible Hamiltonian flows toward its effective version at larger and larger scales. This picture, however, relies on the Hamiltonian uniquely defining the state, and must be given up if we trust the Hamiltonians to encode only the short-range correlations.

The previous discussion is rather idealized, as a lot of work is needed to transform these ideas into concrete methods. This is why most techniques classified under the umbrella of RG are very model-specific and hence of little use for us.

However, the study of quantum phase transitions (i.e. phase transitions at zero temperatures, where all fluctuations are quantum instead of thermal) has motivated the appearance of much more generic numerical methods based on an {\em ansatz}, i.e., a parameterization of a manageably small set of states which happen to nearly contain many of the interesting physical states. For instance, the very successful density-matrix renormalization group (DMRG) can be understood as an optimization method within the matrix product state (MPS) ansatz. Classically, an MPS can be thought of as a type of stochastic finite-state machine (although another classical version has been proposed~\cite{temme10}). This, however, only works on one-dimensional systems, at least in the quantum setting. In addition, the connection with the RG is somewhat weak (even though it is faithful to a proposal made in the seminal paper by Wilson on the renormalization group~\cite{wilson83}).

Here, we will consider instead a more recent proposal which can be formulated in any dimension, and which more directly incorporates the idea of a multiscale description of the system.

\section{Stochastic version of MERA}

The multiscale entanglement renormalization ansatz (MERA) is an efficient parametrization of a certain subset of {\em quantum states} (which one may think of as probability distributions but where the probabilities take complex values). It is numerically efficient in two ways which are important for its use in physics. Firstly, the number of parameters involved typically grows linearly with the number $N$ of sites. Secondly, marginal distributions over a constant number of sites (not necessarily contiguous) can be explicitly computed in a time of order $\log N$. 

Instead of describing the full quantum MERA, I will only introduce a classical version of it. Since all entanglement (a.k.a. quantum correlations) is replaced by classical correlations, let us call it the correlation renormalization ansatz (CORA). 

For simplicity, I will describe the model as it would apply to data living on a square $d$-dimensional lattice, such as a time series for $d=1$ or an image for $d=2$, and assume a renormalization model where the coarse-graining is done by grouping blocks of $2\times 2 \times \dots = 2^d$ neighboring sites. Generalization to other lattices or other coarse-graining scheme are rather straightforward. 

Instead of starting with an algorithm per se, we start with a ``knowledge representation'', that is, a parameterization of a certain set of probability distributions on the observed data. Provided that this parameterization can be stored efficiently, and that the cost function for the learning process be computable efficiently, then an algorithm can be devised to minimize the cost function.

We call $\Gamma_0$ the original square lattice on which the data lives, and $\Gamma_{j}$, $j=1,2,\dots$ the successive coarse-graining of that lattice. Each site of $\Gamma_{j}$ must be thought of as representing a block of $2^d$ sites of $\Gamma_{j-1}$, in the sense that the neighborhood relations between the sites of $\Gamma_{j}$ represent that of the corresponding blocks on $\Gamma_{j-1}$. 

Since the number of sites of $\Gamma_{j}$ is a constant fraction of that of $\Gamma_{j-1}$ ($1/2^d$ in this case), the total number of coarse-graining steps is logarithmic in the total size $N$ of $\Gamma_0$, i.e. in the total number of variables. 

By a state on $\Gamma_j$, we refer to a joint probability distribution for the values of the variables associated with each site of $\Gamma_j$. We denote the set of these distributions by $S(\Gamma_j)$. The idea is to represent a state $\mu \in S(\Gamma_0)$ through a sequence of stochastic maps 
\[
\pi_j: S(\Gamma_j) \rightarrow S(\Gamma_{j-1})
\]
as
\[
\mu = \pi_1\circ\pi_2\circ\dots\circ\pi_{j_{\max}}(e),
\]
where $\circ$ denotes composition, and 
$e$ is some fiducial state on the coarsest lattice $\Gamma_{j_{\max}}$, e.g., the uniform distribution.
In the original MERA, these maps $\pi_j$ are {\em isometric} operator on a Hilbert space, which is the quantum equivalent of stochastic maps, in the sense that they are the most general maps preserving the basic properties of the states.  

One may think of these $\pi_j$ as the {\rm inverse} of the coarse-graining channels defining the renormalization group. Indeed, instead of destroying information by mapping a precise lattice to a coarser one with fewer sites, it introduces information by creating additional sites, hence constructing the full state scale by scale. In this sense, MERA is a ``generative'' interpretation of the RG. 

\begin{figure}
\begin{center}
\includegraphics[width=0.8\linewidth]{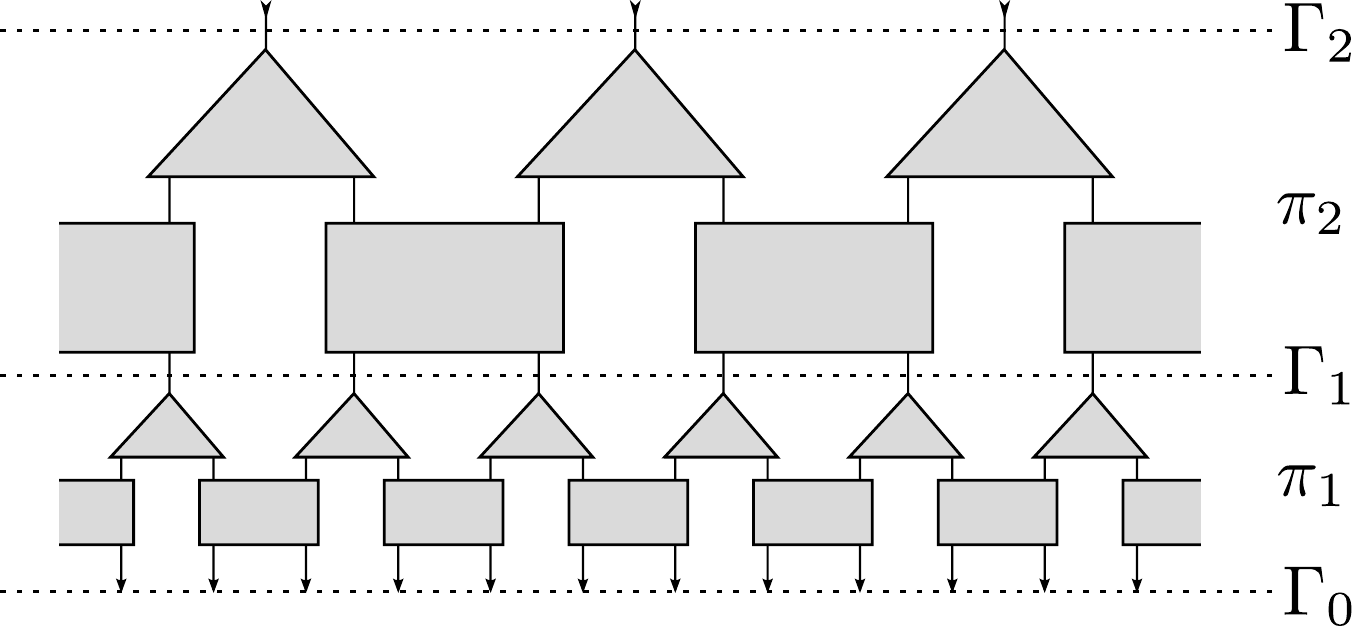}
\caption{Example of a MERA for a one-dimensional lattice, where only two layers are represented. The boxes and triangles are arbitrary isometric maps between Hilbert spaces represented by the vertical lines. We obtain a CORA by replacing the boxes and triangles by stochastic maps between random variables. This diagram can then be read as an operational recipe to produce the state on the lattice $\Gamma_0$, from that of $\Gamma_2$: each vertical line represents some data, or random variable, and a box or triangle is a specific stochastic map which must be applied to the joint probability distribution of all its input(s) (the lines coming in from above) in order to produce the joint distribution of its outputs (represented by the lines coming out below it). This can be equivalently represented by the Bayesian network of Figure~\ref{bayes}. 
The parameters of the model are the components of all the local stochastic maps (boxes and triangles), together with the initial state of the last layer $\Gamma_{j_{\max}}$. 
}
\label{fig1}
\end{center}
\end{figure}

\begin{figure}
\begin{center}
\includegraphics[width=0.8\linewidth]{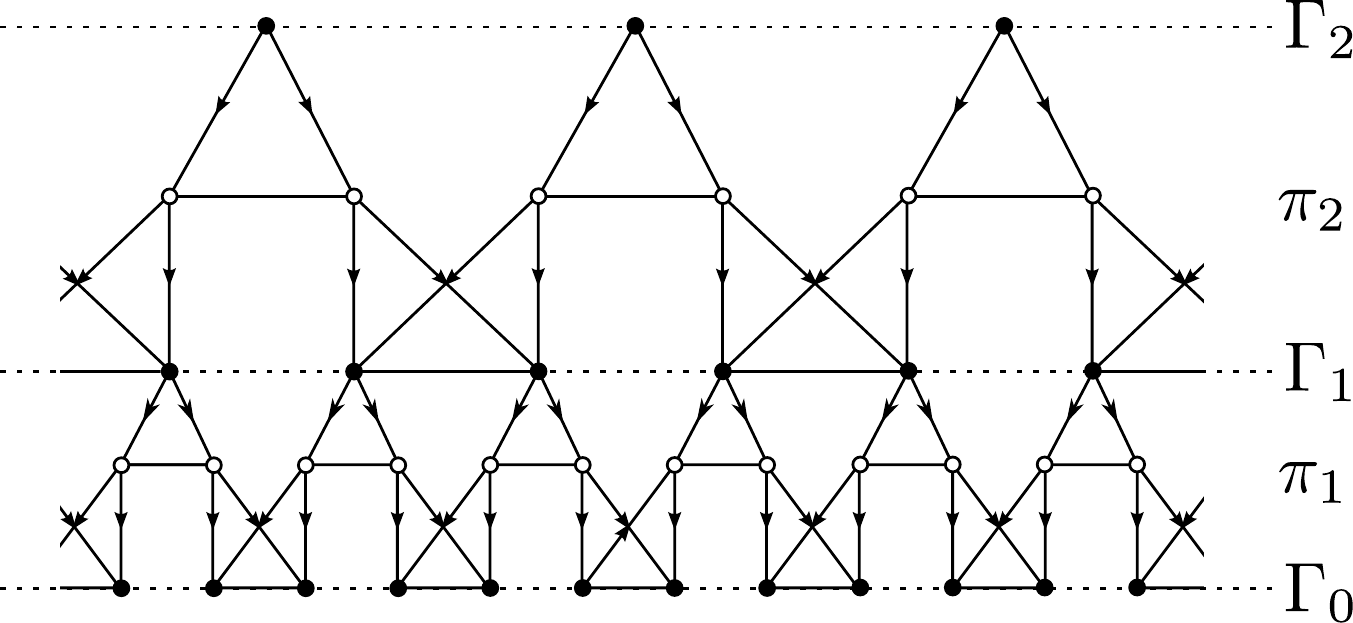}
\caption{Example of Figure~\ref{fig1} represented as a Bayesian network (only two layers are represented). The bottom nodes are observed. Note that the graph is truncated, as the nodes of $\Gamma_2$ must be linked to the next layer which is not represented, as well as to each other, in the same manner as the two layers below it.}
\label{bayes}
\end{center}
\end{figure}

In the original MERA, these maps are specifically required to be implementable by local maps (stochastic maps in the classical case or isometries in the quantum case), such as represented in Figures~\ref{fig1} and~\ref{bayes} for a one-dimensional lattice. Such a given structure being fixed, the variational parameters left to be learned are the components of each local map.

The required decomposability of the maps $\pi_j$ into local operations renders the number of variational parameters linear in the lattice size. 
Apart from this, however, the particular way in which the local maps are combined is not essential, except for the causal properties of the complete maps $\pi_j$ allowed: 
Firstly, $\pi_j: S(\Gamma_j) \rightarrow S(\Gamma_{j-1})$ must be such that, through it, each site of $\Gamma_j$ can only {\em causally influence} the block of sites of $\Gamma_{j-1}$ that it {\em represents}, or their neighbors as well, up to a constant distance. In Figure~\ref{fig1} for instance, it is apparent that a given site can only influence the values of the block of two sites that it represents, immediately below it, together with their nearest neighbors. 

The concept of causality that is used is defined as follows: for a given stochastic matrix with several input and output variables, we say that the $i$th input {\em cannot} causally influence the $j$th output if the variable $i$ is not required for the calculation of the marginal over $j$, no matter what the complete input is. That is, the marginal over $j$ can be computed purely from the marginal over all original input variables but the $i$th.  

Secondly, something which is also apparent in Figure~\ref{fig1}, is the fact that $\pi_j$ cannot introduce any correlations between far-away sites if its input is uncorrelated, where ``far-away'' means that these output sites cannot be influenced by the same input site, in terms of the first causality condition. 

It is in this sense that each layer $\pi_j$ can only create, and therefore represent, correlations at a given scale.

\section{Learning CORA}

\begin{figure}
\begin{center}
\includegraphics[width=0.8\linewidth]{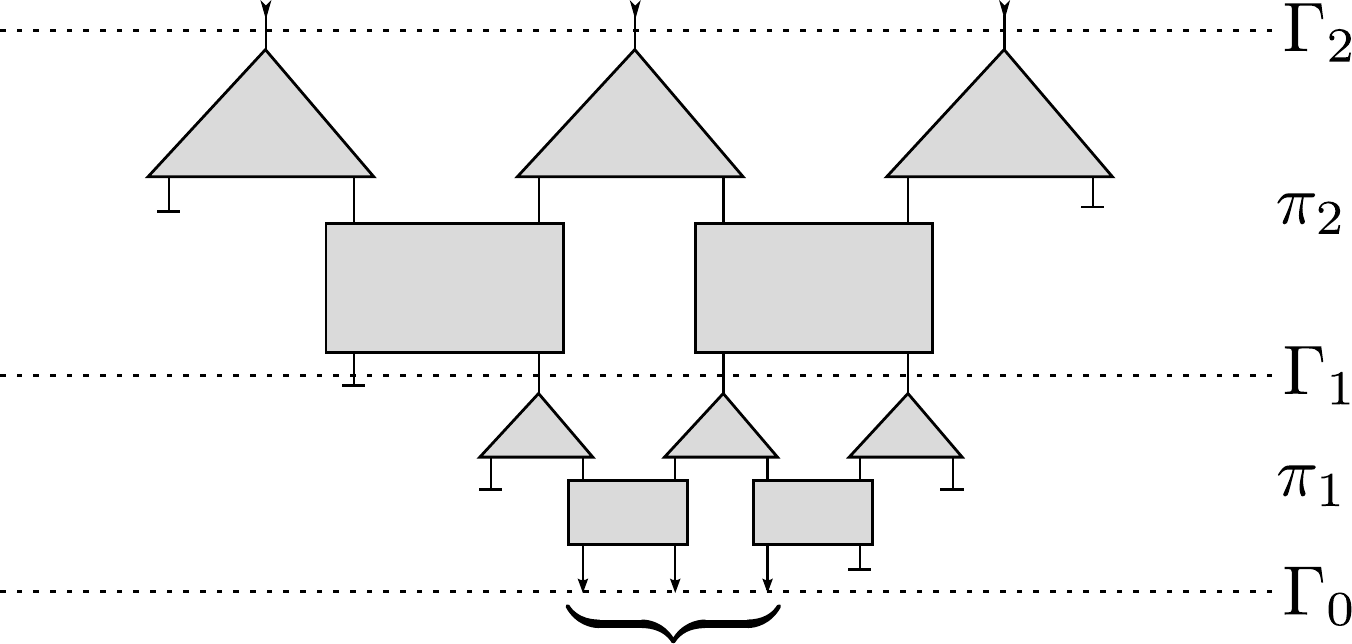}
\caption{Stochastic maps involved in the last two steps of the computation of the marginal state on $3$ consecutive output sites. The lines ending abruptly indicates that the corresponding variable is summed over. The ``past'' of any region of $\Gamma_0$ of size $L$ always involves just $3$ sites before level $\Gamma_{[\log_2(L)]}$.}
\label{causal}
\end{center}
\end{figure}

The causal properties inherent in the definition of MERA/CORA imply that a marginal over any finite group of $L$ sites can be computed (explicitly, i.e., without sampling) in a time of order $e^L \log(N)$. Indeed, due to the particular causal structure of the maps $\pi_j$, the {\em past} of any set of sites of $\Gamma_j$, namely those sites of $\Gamma_{j+1}$ on which their values depend explicitly through $\pi_j$, always ends up involving a constant number of sites independent of $N$ (and generally manageably small). This is illustrated in Figure~\ref{causal}.

In the quantum physical setting for which MERA was introduced, the state that we want to represent is not defined by samples, but instead by a Hamiltonian, or energy function, that it minimizes, i.e., the cost function itself. 
Most often, the Hamiltonians considered are {\em local}, which implies that the evaluation of their expectation only requires the use of marginal states over small clusters of neighboring sites.
Therefore the cost function can be evaluated efficiently and exactly. 

Such a procedure can be adapted to a situation where, instead of being handed the Hamiltonian, we are given samples from the unknown distribution: the training data. In physics, this situation presents itself when an experimentalists wants to reconstruct a state that he has access to only through experiments (which yields the samples), a process which is called ``state tomography''. This is, of course, no different from a typical learning task. 

A difference, however, is that in physics one is often interested in states which minimize some unknown {\em local} Hamiltonian. This assumptions implies that it is sufficient to work with the efficiently computable marginals over small groups of neighboring sites (so called {\em reduced states} in quantum theory). Therefore, we can set the parameters of the channels $\pi_j$ by maximizing the probabilities that these reduced states assign to the training data. In the quantum context, this was proposed in Ref.~\cite{landoncardinal12}.
For CORA, the cost function can be calculated in the same way, but the optimization procedure must be slightly modified to account for the fact that the constraints on the stochastic maps components are different from the quantum MERA.  

\subsection{States not determined by local laws}

It is clear, however, that a MERA, or CORA, can represent a much vaster class of states, in particular states whose long-range correlation structure is not determined by the nature of the short range correlations, such as one would expect for natural or artificial images for instance. Indeed, the components of the stochastic maps $\pi_j$ can be modified directly at any scale $j$ with minimal effect on the short range correlations. 

Let us therefore consider learning in the more general situation. 
Mirroring our discussion of section~\ref{RG}, we can still use the efficiently computable local reduced states in order to optimize the component of the {\em first} channel $\pi_1$, since it cannot anyway encode information about longer ranged correlations. We say that we are working with {\em training length} $\epsilon_0$, ignoring all correlations in the data at larger lengthscales. 

This can be done in two different ways. We may simply optimize $\pi_1$ by locally comparing the training data to $\pi_1(e)$ where $e$ is the uniform distribution, or any other prior (method 1), or we can already optimize all layers simultaneously as would be done for a tomography problem, focusing only on short-range correlation (method 2). This would have the advantage of already making long range---or deep---deductions from the data. For instance, if all neighboring pixels were the same color, this would already deduce that the whole image is uniform.

In a second pass, we need to compare the data to marginals generated over the new training length $\epsilon_1$ associated with the next channel $\pi_2$. We do not want to do this by comparing the data to the full candidate state generated at the level of $\Gamma_0$, as the number of sites involved would grow by a factor $2^d$ with a bad exponential increase in complexity. Instead, we want to directly compare the reduced state at level $\Gamma_1$ to a coarse-grained (i.e., ``renormalized'') version of the training data at that level. 

Performing this step is where the difficulties lie, and could likely not be done quantum mechanically. Classically, however, it may be possible to use sampling in this step. 
This coarse-graining must be done by the Bayesian {\em inverse} of $\pi_1$, with prior $e$ (method 1), or a prior generated by the previous layers (method 2), which is possible because we really only need the prior over scale $\epsilon_1$. 
An example of the type of Bayesian networks involved is represented in Figure~\ref{bayes}.

This process can be iterated until all $\mathcal O(\log N)$ layers/channels have been trained, and the training length covers the whole system.
Further training passes can then be done in the same way in order to exploit the better priors generated by the already trained layers.
 
\section*{Acknowledgments}
Helpful discussions with Tobias Osborne are gratefully acknowledged. 
This work was supported by the cluster of excellence EXC 201 ``Quantum Engineering and Space-Time Research''.

\bibliography{learning13}{}
\bibliographystyle{unsrt}

\end{document}